\documentclass[prb,amssymb,aps]{revtex4}
\usepackage{graphicx}
\usepackage{amsmath,amssymb}

\begin{document}
\draft
\title{Comment on "Search for the Standard Model Higgs Boson
in the Missing Energy and Acoplanar b-Jet Topology at $\sqrt{s} = 1.96$ TeV"
}
\author{E. Donth}
\address{Institut f\"{u}r Physik, Universit\"{a}t Halle, D-06099 Halle (Saale), Germany\\
E-mail: donth@physik.uni-halle.de}

\begin{abstract}
A DO paper of 2008 is exemplarily commented to demonstrate the general risk that the development of a new paradigm for astroparticle physics can be hindered by the work of large collaborations. This paper allows sharp formulations because only the simplest Higgs variant is treated.

{\bf PACS numbers:} 01.10 Hx, 13.85 Rm, 14.80 Bn
\end{abstract}

\maketitle

The research paper \cite{Zitat1} of a larger collaboration is exemplarily criticized with respect
to consequences that often follow from the organization of
communicability inside and between large collaborations,
and from the publication in leading journals of the physical
community for particle physics and cosmology. The
paper from the DO collaboration has 519 authors from 82
groups in 35 large societies worldwide. The publication
and readability necessarily requires the communicability between
several $10^{4}$ scientists. It is shown that the
development of a new Kuhnian paradigm - which must not
be excluded today - is quantitatively suppressed by such
papers.

Suppose that a small-step method is necessary to maintain communicability:
only one or a very few concepts are allowed to be changed. As a specific example, let there be a step succession: SM $\rightarrow$ (y/n)Higgs $\rightarrow$
( y/ n ) SUSY $\rightarrow$ (y / n ) independent quarks $\rightarrow$ ... .
[ For Higgs, y = yes means, according to Nambu \cite{Zitat2} , that the "mass
of a particle is a self-energy due to interaction", and
n = no possibly means, alternatively, Pauli's vision
that a quantum theory should also define its numerical
parameters; e.g. via some kind of renormalization accompanying any experiment, as slightly suggested for charge
and mass. SM = standard model, SUSY = supersymmetry, and
independent quarks are related to a hypothetical quark gluon plasma
without three-legged animals.] Assume that the new paradigm - right
or wrong - contains 20 new concepts. Then 20 quite independent y/n decisions are hidden.
The probability to arrive at the new paradigm would be of order $P = 2^{-20} \approx 10^{-6}$,
i.e. practically zero.

Alternatively, a new paradigm may be arrived by a neural network method.
The old paradigm is partitioned into 20 remainders that are connected
anew by consistency.
This method is suggestive of the connections between neurons in the human brain and
is able to learn by iterations. An off-site model is invented if the iterations
converge off the succession. To get the model, about 2000
trials seem to be sufficient. This number stems from the
iterations of the 20 remainders, of the $20\cdot19/2 = 190$
connections, and of the many more new relations between
them. One trial costs, by experience, about 10 working
hours. Then the method requires communicability by
the inventor over about 40 years. Nevertheless, the result is neither communicable nor
able to be published in today's community.

The situation becomes even more serious by a sociological
factor from the large and increasing number ($N$) of committed scientific participants. Psychologically, infinite
fluctations ($\alpha < 2$) or even infinite expectations ($\alpha < 1$)
for a gain of reputation (fame) require the treatment of
L\'{e}vy limit distributions. The hierarchical structure in
a L\'{e}vy sum is described in \cite{Zitat3}. Call ($X_{1} + X_{2}$ + ...
$+ X_{N}) / N^{1/\alpha}$ the scaled ordered L\'{e}vy sum, $X_{i}$ (i = 1, 2, ..., $N$)
their random variables for the gain, and $\alpha$ the L\'{e}vy
exponent. The amplification factor $N^{1/\alpha}$ is necessary
for scaling and for obtaining a nontrivial limit distribution. Assume the ratio
of new to old numbers is $N'/ N''$
= 100. Then, from hierarchy arguments, the ratio of powers
of the corresponding communities is $1/P' = (N'/N'' )^{1/\alpha}
= 100^{1/\alpha}$. [The fame may be suggested to be larger and
in their time, if there is no alternative.] Gauss
($\alpha = 2$) is not critical, $P'$ = 1/10; Cauchy ($\alpha = 1$) at
the borderline to infinite expectation is $P'$ = 1/100,
and greed for fame ($\alpha = 1/2$) is $P'$ = 1/10000: large
power, sharp hierarchies, and preponderant components in
the communities.

The total probability to get a new paradigm amounts
to $P$(tot) $= PP'= 10^{-10} \approx 0$ for $\alpha = 1/2$. This result excludes
definitely that an off-site model - if existing - can
be detected via communicability and can be publicated today
or in the near future. [The \cite{Zitat1} experiments are very
useful, but the parameter handling may be risky without
Higgs.] This communication difficulty may leave physics in a trap, by overstretching
and exhaustion of the old paradigm. The knowledge of a new paradigm, however,
is necessary for a Kuhnian revolution: "Competition
between segments of the scientific community is the only historical
process that ever results in the ... [change
of paradigms]" \cite{Zitat4}. The way out from the hard dilemma is
recovery of liveliness. We need a method offering a fast way to the acceptance of competing segments.
The task concerns two institutions: the publication
journals and the scientific organizers of the collaborations. Serious alternative
suggestions should become
publishable, perhaps under the category of "paradigmatic
off-site models", before belatedly a certain number of noes
in the succession becomes obvious; and should not be
forbidden in the collaborations, but at least partly be
supported. The problem is the growing numbers: of new
concepts for a new paradigm and of committed scientists.

\end{document}